\documentclass{article}
\usepackage{amsmath,amsfonts,amscd} 
\usepackage{bm}
\usepackage{graphicx}
\def\bra{\langle} 
\def\ket{\rangle} 
 
\newcommand{\C}{\mathbb{C}}

\newcommand{\Q}{\mathbb{Q}}
\newcommand{\R}{\mathbb{R}}

\newcommand{\Z}{\mathbb{Z}}

\newcommand{\cH}{\mathcal{H}} 
 
\newcommand{\cQ}{\mathcal{Q}} 
 
\newcommand{\cZ}{\mathcal{Z}} 
\def\lr{{L^2(\R)}} 
\def\l1r{{L^1(\R)}}

\def\Qp{\cQ_p} 
\def\Zp{\cZ_p}

\def\eg{{\sl e.g. }} 
\def\etc{{\sl etc.}}

\begin{document}
\title{$p$-Adic wavelet transform and quantum physics\thanks{Talk at 1st 
Int. Conf. on $p$-Adic Mathematical Physics, Moscow, Oct 1-5, 2003.}}
\author{M.V.Altaisky\\ Joint Institute for Nuclear Research, 
Dubna, 141980, Russia; \\ and Space Research Institute RAS, Profsoyuznaya 
84/32, Moscow,\\ 117997, Russia, e-mail: altaisky@mx.iki.rssi.ru}
\date{}
\maketitle
\begin{abstract}
$p$-Adic wavelet transform is considered as a possible tool for the 
description of hierarchic quantum systems. 
\end{abstract}
\section{Introduction}
Wavelet transform (WT) (see any general textbook \cite{Daub10,Mallat:book} for 
the detailed review of the subject), as a generalization of windowed 
Fourier transform is widely used in various branches of physics, 
computing and image processing. Its tremendous success in localizing the 
singularities of
fields and signals is mainly due to its mutual locality in coordinate
and momentum: a local disturbance of
wavelet coefficients affects the reconstructed image only {\it locally},
without spoiling the whole picture.

Formally the wavelet decomposition of a vector $\nu$ in a Hilbert space $\cH$ 
is 
a decomposition with respect to the square-integrable representation $U(g)$ 
of the affine group 
\begin{equation}
G:x'=ax+b, \quad a,b \in \R, a\ne0, \label{ag}
\end{equation}
 acting transitively on $\cH$ \cite{GMP1985}, written as follows: 
$$
|\nu\ket = \int_G U(g) |\psi\ket d\mu_L(g) \bra \psi | U(g) |\nu \ket, 
$$
where $d\mu_L(g) = \frac{dadb}{a^2}$ is a left-invariant measure on $G$ and  
$\psi \in \cH$ is an appropriately chosen vector, called a basic wavelet.

In practical applications, $\cH$ usually is the space of square-integrable 
functions $\lr$, and the basic wavelet $\psi$ is a certain compactly supported 
function in it. The power of the wavelet transform is due to the fact that 
the analyzing window function 
$[U(a,b)\psi](x) = |a|^{-1/2} \psi\left(\frac{x-b}{a}\right)$ has the width 
varying with the scale parameter $a$, and thus works as a microscope 
that locally analyzes the details of a given scale. 

Bearing in mind the application to the functions of $p$-adic argument 
rather than $\lr$, we modify the definitions of direct and inverse WT in 
$\l1r$-norm as follows: 
\begin{eqnarray}
 (W_\psi f)(a,b) &=& \int \frac{1}{a} \bar \psi 
\left(\frac{x-b}{a} \right) f(x) dx \label{dwt} \\
f(x) &=& \frac{2}{C_\psi} \int_0^\infty \frac{da}{a^2} \int db (W_\psi f)(a,b) 
\psi\left(\frac{x-b}{a}\right) \label{iwt},
\end{eqnarray}  
where 
$$C_\psi = \| \psi \|^{-2} \int d\mu_L(g) |\bra\psi|U(g)|\psi\ket|^2.$$
For the case of affine group \eqref{ag} 
$C_\psi = \int \frac{|\tilde\psi(\omega)|^2}{|\omega|}d\omega$, where tilde 
means the Fourier transform.

Before defining wavelet transform in $\Qp$, it is worth noting that 
both wavelet transform and $p$-adic numbers provide hierarchic description 
of reality. 
%%%%%%%%%%%%%%%%%%%% this part was excluded in journal version %%%%%%%%%%
Say, $p$-adic integers, defined by \eqref{pint} below, 
for $p\!=\!2$ describe a point of the unit interval $[0,1)\!\in\!\R$ in 
terms of the embedded set of its vicinities, halved in size at each step. 
To some extent, a $p$-adic integer \eqref{pint} labels an analog  
point in a $(d\!=p\!-\!2)$-dimensional simplex partitioned in a self-similar 
way into $p$ parts on each iteration,  see Fig.~\ref{trg2:pic} for two 
dimensional example. Since the 
$d$-dimensional sphere $S^d$ is homeomorphic to the boundary of $(d\!+\!1)$ 
dimensional simplex, it is possible to label the points of $S^d$ using the 
coordinates on that simplex.
%%%%%%%%%%%%%%%%%%%%%%%%%%%%%%%%%%%%%%%%%%%%%%%%%%%%%%%%%%%%%%%%%%%%%%%%

The remainder of this paper is organized as follows. In {\em Section 2} 
we review the basic facts on $p$-adic numbers, their geometrical 
interpretation and  consider a few constructions of $p$-adic wavelet 
transform. An implementation of the $p$-adic WT with the Haar wavelet is 
presented. In {\em Section 3}, we define $p$-adic WT with Haar wavelet. 
{\em Section 4} presents the Hilbert space of states 
of hierarchic quantum systems and considers possible links to $p$-adic wavelet 
transform.  A generalization of quantum computation 
on qubit states is extended in {\em Section 5} to quantum 
systems constructed from elementary units with $p\!>\!2$ states. 

\section{p-Adic wavelet transform}
Prior to defining $p$-adic wavelet transform, let us review the 
basic facts about $p$-adic numbers. The results of any physical measurement 
can be expressed in terms of rational numbers $\R$. The construction of a
theoretical model, first of all a description in terms of differential
equations, ultimately requires an extension of this field. The first
extension
(made by Dedekind, viz. equivalence classes) means the incorporation of
rational numbers. It is a completion of the field $\Q$ with respect to
the standard norm $\Q \stackrel{|\cdot|}{\rightarrow} \R$. However, this 
completion
 does not exhaust all possibilities. It is also possible to extend
the field $\Q$ using the {\it p-adic norm} $|\cdot |_p $
(to be explained below):
\begin{equation}
\R \stackrel{|\cdot|}{\leftarrow} \Q
\stackrel{|\cdot|_p}{\rightarrow} \Qp.   \label{ost}
\end{equation}
This completion is called the field of $p$-adic numbers $\Qp$.
No other extensions of $\Q$ except those two exist due to the
Ostrovski theorem \cite{Ost1918,BS1966}.

The $p$-adic norm $|\cdot |_p$ is defined as follows.
Any nonzero rational number $x \in \Q$
can be uniquely written in the form
\begin{equation}
x = \frac{m}{n}p^\gamma,  \label{pdec}
\end{equation}
where integers $m$ and $n$ are not divisible by the prime integer $p\ne1$,
and $\gamma$ is an integer.  The decomposition (\ref{pdec}) provides
a possibility to supply the field $\Q$ with the
norm
\begin{equation}
|x|_p = {\left| \frac{m}{n}p^\gamma \right|}_p \label{pnorm}
\stackrel{def}{=} p^{-\gamma}, \qquad |0|_p \stackrel{def}{=} 0,
\end{equation}
different from the standard one. The algebraic closure of the field
$\Q$ in the norm $|\cdot|_p$ forms the field of $p$-adic numbers
$\Qp$.

Any $p$-adic number can be uniquely written in the form
\begin{equation}
x = \sum_{n=k_{min}}^{\infty} a_n p^n,
\qquad a_n \in \{0,1,\ldots,p-1\}, \quad k_{min} > -\infty.
\end{equation}
It is easy to see that
$|xy|_p = |x|_p|y|_p,$ but $|\cdot|_p$ is stronger than $|\cdot|$:
\begin{equation}
|x+y|_p \le max (|x|_p,|y|_p) \le |x|_p+|y|_p,
\end{equation}
and induces a non-Archimedian metric
\begin{eqnarray}
d(x,y) &:=& |x-y|_p \label{pdist}\\
\nonumber d(x,z) &\le& max (d(x,y),d(y,z)) \le d(x,y) + d(y,z),
\end{eqnarray}
often called an {\em ultrametric} \cite{RV1986}.
With respect to the metric (\ref{pdist}), the $\Qp$ becomes a
complete metric space. The maximal compact subring of $\Qp$
\begin{equation}
\Z_p = \{ x \in \Qp : |x|_p \le 1 \}
\label{pint}
\end{equation}
is referred to as {\em a set of p-adic integers}.
The field $\Qp$ admits a positive Haar measure,
unique up to normalization
\begin{equation}
d(x+a) = dx, \quad
d(cx)  = |c|_p dx, \quad x,a,c \in \Qp.
\label{hmp}
\end{equation}
The normalization is often chosen as
$\int_{\cZ_p} dx \equiv 1$.

The geometry induced by the distance $|x-y|_p$ is quite different
from the Euclidean one: all $p$-adic triangles are equilateral;
two $p$-adic balls may either be one within another or disjoint.

There is no unique definition of differentiation in the field $\Qp$,
but the Fourier transform exists and is used in $p$-adic field
theory to construct the pseudo-differential operator
$$D \phi(x) \to |k|_p \tilde \phi(k).$$
The construction of the $p$-adic Fourier transform is essentially
based on the group structure of the field $\Qp$, viz. the group
of additive characters
\begin{equation}
\chi_p(x) := \exp \bigl( 2\pi\imath \{ x \}_p \bigr),
\quad \chi_p(a+b) = \chi_p(a)  \chi_p(b),
\label{char}
\end{equation}
(where $\{ x \}_p$ denotes the {\it fractional part} of $x$:
$\{ x \}_p = a_{min} p^{k_{min}} + \ldots + a_{-1} p^{-1}$),
is used to construct the Fourier transform 
\begin{equation}
\tilde \psi(\xi) = \int_{\Qp} \psi(x) \chi_p(\xi x)dx,\quad
\psi(x) = \int_{\Qp} \tilde \psi(\xi) \chi_p(-\xi x)d\xi.
\end{equation}
The $n$-dimensional generalization is straightforward. 
%\begin{eqnarray*}
$$
\Qp\to\Qp^n, \quad
x\to(x_1,\ldots,x_n), \quad
\xi\to(\xi_1,\ldots,\xi_n), \quad
\xi x\to(\xi,x)=\sum_i\xi_i x_i.
$$
%\end{eqnarray*}

\begin{figure}[ht]
\centering \includegraphics[width=6cm]{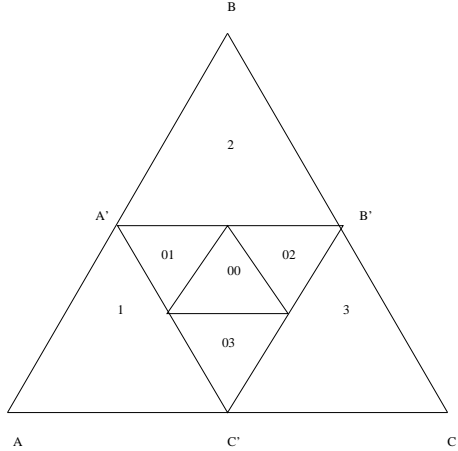}
\caption{The partition of a d-dimensional simplex ($d\!=\!p\!-\!2$) into $p$
 equal
parts.The case of $d\!=\!2$ is shown. If the the unit measure is assigned to 
the initial simplex, the volume of each subpart coincides with the 
norm of the $p$-adic number it is labeled by.}
\label{trg2:pic}
\end{figure}

Now, let us show how $p$-adic coordinates can be used to label the 
points of ($d\!=\!p\!-\!2$)-dimensional manifolds. Let us consider a 
($d\!=\!p\!-\!2$)-dimensional simplex, a triangle for $d\!=\!2$ is shown 
in Fig.~\ref{trg2:pic}. 
We can divide our 2d simplex, shown in Fig.~\ref{trg2:pic}, into 
into $p\!=\!4$ equal equilateral triangles and label them by 0,1,2,3.
The initial triangle is therefore subdivided into $p\!=\!d\!+\!2$ equal
parts; since the same procedure can be also applied to its parts
we can move down {\sl ad infinum} and label these parts as 
$a_0,a_1,\ldots$,
where $0\le a_i < p$. If we ascribe a measure $\mu(T_0)=1$ to the
initial triangle, when at the first stage we have $p$ triangles of
the measure $1/p$ each, at the second stage $p^2$ triangles of the measure
$1/p^2$ and so on. To meet this natural definition of the measure, we can
label the sequences $\{ a_i \}$ by $p$-adic integers. Generally, a 
$d$-dimensional simplex can be divided into $p=d+1$ equal parts, 
but to some extent we can consider $p$-adic balls as vicinities of 
points in a compact ultrametric space with a geometry similar to $S^d$ 
partitioning shown in Fig.~\ref{trg2:pic}.

Having defined the integration in $\Qp$ by \eqref{hmp}, we can formally 
generalize the continuous wavelet transform to the affine group acting 
in $\Qp$ \cite{AltP1997}:
\begin{equation}
G:x' = ax + b, \quad a,x,b \in \Qp, \quad a\ne0. \label{ap}
\end{equation}
As it follows from the group multiplication law
$$ (a,b)(a',b') = (aa',ab'+b),$$
the left invariant measure on $p$--adic affine group \eqref{ap},
up to normalization, has the form
\begin{equation}
d\mu_L(a,b) = \frac{dadb}{|a|_p^2}, \label{muap}
\end{equation}
which provides
$$ d\mu_L(aa',ab'+b) = \frac{|a|_p da'|a|_p db'}{|aa'|_p^2}
                     = \frac{da'db'}{|a'|_p^2}. $$
To construct a counterpart wavelet transform (\ref{dwt},\ref{iwt}) 
for the $p$-adic affine group \eqref{ap},
we use an obvious definition scalar product of complex-valued functions
of $p$-adic arguments,
$$ \bra f,g \ket = \int_{\Qp} \bar f(x) g(x) dx, $$
which provides a functional norm
$\| f \|^2 \equiv \bra f,f \ket$.
Thus the direct generalization of (\ref{dwt},\ref{iwt}) takes the
form
\begin{eqnarray}
 (W_\psi f)_{\Qp} (a,b) &=& \int_{\Qp} \frac{1}{a} \bar \psi 
\left(\frac{x-b}{a} \right) f(x) dx \label{dwtp} \\
f(x) &=& \frac{1}{C_\psi} \int_{\Qp^*\times\Qp}  
\psi\left(\frac{x-b}{a}\right)(W_\psi f)_{\Qp} (a,b)\frac{dadb}{|a|_p^2}
 \label{iwtp},
\end{eqnarray}
where $\Qp^*\equiv \Qp\setminus0$.

An analog of the complex-valued Morlet wavelet have been constructed by 
S.Kozyrev \cite{Koz2002}.This complex valued wavelet is constructed 
from additive character \eqref{char} of the field $\Qp$ used in 
$p$-adic Fourier transform, and has the form 
\begin{equation}
\psi(x) = \chi(p^{-1}x) \Omega(|x|_p),\quad x \in \Qp,
\label{koz}
\end{equation} 
where $\Omega(\cdot)$ is characteristic function of the unit 
interval in $\R$. 
The Kozyrev wavelets \eqref{koz} are eigenfunctions of the 
Vladimirov p-adic pseudoderivative \cite{VVZ1994}:
\begin{equation}
D^\alpha \psi(x) = p^\alpha \psi(x), \quad 
D^\alpha f(x) := \frac{p^\alpha-1}{1-p^{-1-\alpha}} \int_{\Qp}
\frac{f(x)-f(y)}{|x-y|_p^{1+\alpha}} dy,
\end{equation} 
where $dy$ is the Haar measure on $\Qp$. 

\section{p-Adic wavelet transform with Haar wavelet}
Concerning the discrete implementation of the $p$-adic wavelet 
transform, the most evident implementation is that with Haar wavelet.
Haar wavelet is the simplest case of the basic wavelet  
\begin{equation}
h(x) = \begin{cases}
       1 :& 0 < x < 1/2 \\
       -1:& 1/2 \le x < 1 \\ 
        0:& {\rm elsewhere}. \cr
       \end{cases}
\end{equation} 
Using the scale factor $a\!=\!2$, we easily obtain the discrete representation 
of the affine group ($\l1r$ norm is used) 
$$
h^j_k(x) = 2^{-j} h(2^{-j}x-k)$$
and the wavelet coefficients 
$$
d^j_k[f] = \int 2^{-j} h(2^{-j}x-k) f(x) dx.$$
Then, forward and backward discrete wavelet transform with 
the Haar wavelet can be then easily evaluated using the Laplacian 
pyramidal scheme
$$
\begin{array}{lllllll}
s^0&\to     & s^1 &\to      & s^2 & \to     & \ldots \\
   &\searrow&     &\searrow &     &\searrow &        \\
   &        & d^1 &         & d^2 &         & \ldots
\end{array},
$$   
where $s^0$ is the initial data set, and two projections 
\begin{equation}
s_k^{j+1} = \frac{s^j_{2k}+s^j_{2k+1}}{2}, \quad 
d_k^{j+1} = \frac{s^j_{2k}-s^j_{2k+1}}{2}
\label{dhwt}
\end{equation}  
are evaluated on each step; 
$s^{j+1}$ is evidently the ``blurred'' version of the 
previous level coefficients $s^j$, with $d^{j+1}$ being the details 
blurred out.
The reconstruction from the Haar wavelet decomposition \eqref{dhwt} is
\begin{equation}
s^j_{2k+1} = s^{j+1}_k - d^{j+1}_k, \quad 
s^j_{2k}   = s^{j+1}_k + d^{j+1}_k.
\label{ihwt}
\end{equation}

Similar fast algorithms can be constructed for a $p$-adic Haar wavelet. 
We define an analog of the Haar wavelet in $\Qp$ as 
\begin{equation}
h(x) = \begin{cases}
       1 & |x|_p < 1 \\
       ``-1'' & |x|_p \ge 1
       \end{cases}, 
\quad x \in \Qp
\end{equation} 
where ``-1'' means $(-1\!\in\!\R)$ or $\bar1\!=\! (p-1) + (p-1)p + (p-1) p^2 + \ldots$,
depending on whether $h$ is a complex-valued $h:\Qp\to\C$ or a $p$-adic-valued 
function $h:\Qp \to \Qp$ . In the latter case,  
the pyramidal algorithm (\ref{dhwt},\ref{ihwt}) for the Haar wavelet 
is exactly reproduced in $p$-adic arithmetics 
\begin{equation}
s_k^{j+1} = s_{2k+1}^j + s_{2k}^j, \quad
d_k^{j+1} = s_{2k+1}^j - s_{2k}^j \equiv s_{2k+1}^j + \bar s_{2k}^j, 
\end{equation}
with the reconstruction formulae  
\begin{equation}
s^j_{2k+1} = \frac{s^{j+1}_k + d^{j+1}_k}{2}, \quad 
s^j_{2k} = \frac{s^{j+1}_k + \bar d^{j+1}_k}{2}.
\end{equation}  
\section{Hierarchic states of quantum systems}
The quantum mechanical description based on the Schr\"odinger equation is 
formally applicable to any physical system from elementary particles 
to macroscopic bodies. However, it is technically impossible to account 
for each electron wave function in a living cell or in a microprocessor, or 
even in a big cluster of atoms. The methods of 
quantum statistical mechanics are of little use here, for most quantum 
systems of practical interest are far from statistical equilibrium.

At the same time, the problem of quantum description of mesoscopic objects, 
such as clusters of atoms and living cells, are becoming of steady practical 
interest for high-performance computing and biotechnology. Fortunately, there 
is a way to construct the wave functions of composite objects without taking 
the direct product of all component wave functions. As we always see, the 
Nature has clearly manifested its hierarchic structure: an electron is part 
of atom, an atom is part of a molecule \etc. 
This suggests that instead of taking the direct product of the 
component wave functions, we can represent the wave function of a composite 
system in a hierarchic form, successively taking into account the 
states of all systems our system is part of.

\begin{figure}[th]
\centering \includegraphics[width=3in]{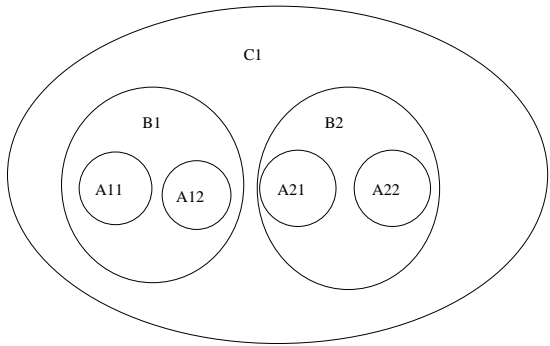}
\vspace*{8pt}
\caption{The structure of binary hierarchic system}
\label{hier1:pic}
\end{figure}

This suggests an idea of the hierarchic wave function \cite{AltIJQI2003}. 
Let the system $C_1$ consist of two subsystems $B_1$ and $B_2$, each of 
those in turn consists of two subsystems, \etc, see Fig.~\ref{hier1:pic}. 
To describe the binary hierarchic system shown in Fig.~\ref{hier1:pic}, we 
need a collection of wave functions 
\begin{equation}
\Psi = \{ \psi_{C_1},  \{ \psi_{C_1B_1},\psi_{C_1B_2}\}, 
\{ \psi_{C_1B_1A_{11}},\psi_{C_1B_1A_{12}}, \psi_{C_1B_2A_{21}},\psi_{C_1B_2A_{22}}\}
 \},
\label{eq1}
\end{equation}
where $\psi_{C_1}$ is the wave function of the whole (labeled by $C_1$), 
and $\psi_{C_1B_1}$ is the wave function of a {\sl 
component} $B_1$ belonging to the entity $C_1$. The subsystem $B_1$ of $C_1$ 
described by the wave function $\psi_{C_1B_1}$ is not the same as a free 
system $B_1$ described by wave function $\psi_{B_1}$; \eg an electron in atom 
is not the same as a free electron. The component wave function, 
that completely determines the state of the subsystem $A_{11}$ is then  
$(\psi_{C_1},\psi_{C_1B_1},\psi_{C_1B_1A_{11}})$.
If required, the density matrix of such a system can be obtained by averaging 
over degrees of freedom of $C_1$ and $B_1$, but not the $B_2$, 
see \cite{AltIJQI2003} for details. 

Evidently, the binary tree structure presented above is a mathematical 
idealization: there is no {\em a priory} reason for the subsystem to 
have exactly the same number of parts as the system it is part of. 
This idealization, however, seems to be a good starting point to 
build a Hilbert space of states of hierarchic system. Having the same 
number of parts at each hierarchy level, and thus being self-similar, 
the presented construction of the hierarchic wave function gives 
us a quantum mechanics on $\Qp$, where $p$ is the number of subparts.   

Let us consider the Hilbert space of hierarchic wave functions 
\cite{AltIJQI2003}. If $\Psi_1$ and $\Psi_2$ are two hierarchic wave 
functions (describing the same type hierarchic objects), then 
the superposition principle requires their linear combination to 
be in the same Hilbert space  
$$\Psi_1,\Psi_2 \in \cH, a,b \in \C \Rightarrow 
\Psi = a\Psi_1+b\Psi_2 \in \cH.$$
For instance, if  
\begin{eqnarray*}
\Psi_1 &=& \{ \psi_{B_1},  \{ \psi_{B_1A_1},\ldots,\psi_{B_1A_N}\},\ldots \}, \\
\Psi_2 &=& \{ \psi_{D_1},  \{ \psi_{D_1C_1},\ldots,\psi_{D_1C_N}\},\ldots \},
\end{eqnarray*}
then their linear combination is 
\begin{equation}
\Psi = a\Psi_1+b\Psi_2 = \{ a\psi_{B_1}\!+\!b\psi_{D_1}, 
                         \{ a\psi_{B_1A_1}\!+\!b\psi_{D_1C_1}, \ldots, 
                            a\psi_{B_1A_N}\!+\!b\psi_{D_1C_N}\}, \ldots\}
\label{multilin}.
\end{equation}
The scalar product is defined componentwise: 
\begin{equation}
\bra \Psi_1|\Psi_2\ket = \bra \psi_{B_1}|\psi_{D_1} \ket 
+ \sum_{i=1}^N \bra \psi_{B_1A_i}|\psi_{D_1C_i} \ket + \ldots.
\label{scprod}
\end{equation}
The norm of the vector in hierarchic space defined by scalar product 
is a sum of norms of all components: 
\begin{equation}
||\Psi||^2 = \bra \Psi_1|\Psi_1\ket = \bra \Psi_{B_1}|\Psi_{B_1}\ket 
                + \sum_{i=1}^N \bra \Psi_{B_1A_i}|\Psi_{B_1A_i}\ket 
                + \ldots.     
\label{mnorm}
\end{equation}
For a self-similar partitioning, if the number ($p$) of subparts 
is the same at all hierarchic levels, we can easily cast the 
properties (\ref{multilin}-\ref{mnorm}) for the wave functions of $p$-adic 
argument $x\in\Zp$. Let 
\begin{equation}
x = a_0 + a_1 p + a_2 p^2 + \ldots, \quad 0 \le a_i < p
\end{equation}
Then, we can define the hierarchic wave function component as 
\begin{equation}
\Psi(x\in\Zp) = \{\psi_{a_0},\psi_{a_0,a_1},\psi_{a_0,a_1,a_2},\ldots\}
\label{pvf}.
\end{equation}
with normalization condition 
\begin{equation}
\int_{\Zp}\bar\Psi(x)\Psi(x) dx = 1.
\label{pvfn}
\end{equation}
At this point we ought to mention the geometric aspect of the 
hierarchic wave function \eqref{pvf}. If $x\!=\!a_0$ labels 
an entity, and $x'\!=\!a_0\!+\!a_1p$ labels the parts of this 
entity, then the Haar measure $dx$ normalized as $\int_{\Zp}dx\!=\!1$ 
provides that the measure the wave function $\psi_{a_0a_1}$ 
is integrated over is exactly $1/p$ of the measure of the entity $a_0$.
This impose a particular geometry to the quantum mechanics of hierarchic 
wave functions $\eqref{pvf}$, which may, or may not, correspond to physical 
reality.

The second quantization on hierarchic states also has its own peculiarities.
To create a state vector \eqref{eq1} of a hierarchic state, we have first 
to create the entity ($C_1$), and only then, we can create its parts 
acting by appropriate creation operators to the state $|C_1\ket$.
\begin{eqnarray}
\nonumber a^+(C_1)|0\ket = |C_1\ket, 
a(C_1)|C_1\ket = |0\ket, a(C_1)|0\ket = 0 |0\ket, \\
a^+(B_i) |C_1\ket = \{|C_1\ket,|C_1B_i\ket\},
a(C_1)\{|C_1\ket,|C_1B_i\ket\} = |B_i\ket.
\label{squant}
\end{eqnarray}
The latter equation \eqref{squant} means the annihilation of 
the entity $C_1$ in a system comprised by components $B_i$: this 
is a decay of $C_1$ into components. 
The possible action of annihilation operator of a part $B_i$ in entity 
$C_1$ 
$$a(B_i)\{|C_1\ket,|C_1B_i\ket\}= |C_1\ket\quad (?)  $$
is questionable. 

If a toy model with the same number of parts at each hierarchy level is 
considered, the second quantization rules \eqref{squant} provide 
a second quantization for the state vectors labeled by $p$-adic numbers. 
The hierarchic state 
vector can be created by sequential action of creation 
operators 
\begin{equation}
|x\ket = |a_0a_1a_2\ldots \ket = \ldots a^+(a_2) a^+(a_1) a^+(a_0)|0\ket,
\end{equation} 
where second quantization $[a(i),a^+(j)]=\delta_{ij}$ is defined on 
the cyclic group $Z_p$ rather than a ring of $p$-adic integers. 
\section{Qubits and $p$-qubits}
Qubit, a quantum analog of a classical bit, is a superposition of 
any two orthogonal quantum states, labeled as ``0'' and ``1'' for definiteness
\begin{equation}
|x\ket = \alpha_0 |0\ket + \alpha_1|1\ket.
\label{qubit}
\end{equation}
Qubit is an elementary unit of quantum information. Strings of qubits 
$$
|x_1,\ldots,x_n\ket \equiv |x_1\ket \otimes\ldots\otimes |x_n\ket
$$ 
are considered as analogues of classical computer registers. The general 
review in quantum information can be found elsewhere, see \eg \cite{nc2000}. 

The idea of qubit, a superposition of two quantum states, comes 
from the fact that it is easy to prepare such states, \eg a spin system 
or a two-level atom. Nevertheless, quantum systems with more than two 
orthogonal states are also known: the Potts model, atoms with 3 and more 
states \etc. Interestingly, it is very likely that the processing of 
genetic information in living cell operates as a quantum information 
processing in mod 4 or mod 5 arithmetics \cite{AFO2003}. 
So, we need a generalization of 
qubit to a superposition of ($p\!>\!2$) orthogonal quantum states. We will 
call it $p$-qubit. 

A particular generalization of this type, a pentabit $p\!=\!5$, has been proposed 
in \cite{AFO2003} 
\begin{equation}
|\phi\ket = \alpha_0 |0\ket + \alpha_1|A\ket +\alpha_2 |C\ket +\alpha_3|T\ket
+\alpha_4|G\ket,
\label{pentabit}
\end{equation}
where $A,C,T,G$ stand for the four nucleotides (adenine, cytosine, thymine, 
guanine), 
and $0$ means the vacuum state, or a gap in a nucleotide sequence. 
The particular correspondence between 
nucleotides and the cyclic group $Z_5$ comes from biochemical properties 
of four nucleotides. Since $A$ and $G$ are purines and $C$ and $T$ are pyrimidines,  
the wave functions of $A$ and $G$ ($C$ and $T$, respectively) should not 
be very 
different from each other \cite{Patel2001}; thus it seems reasonable to use 
representation providing $A+G=0, T+C=0$, as chosen above in \eqref{pentabit}. 

A string of such p-qubits can be subjected to all standard operations of 
quantum computing, including quantum Fourier transform and quantum wavelet 
transform, with the only exception that mod 2 operations will be substituted 
by mod $p$ operations. If a nucleotide sequence is considered as such a 
string ($p=5$), $p$-adic quantum computing may be a biocomputing. 

\paragraph{Computational basis.}
Similar to the traditional binary quantum computing,  
we can define functions and operators acting on $p$-qubits. 
Let 
\begin{equation}
|x\ket \otimes |s\ket , \quad x,s=0,\ldots,p-1
\end{equation}  
be a computational basis. Let $f(x)$ be  
a mapping $Z_p \to Z_p$. The function $f$ can be then represented by 
an operator 
\begin{equation}
U_f |x\ket \otimes |s\ket = |x\ket \otimes |s\oplus f(x)\ket.
\label{operat}
\end{equation}
\paragraph{Walsch-Hadamard transform.}  
For each $p$-qubit $|x\ket$ we can define a Hadamard  transform  
\begin{equation}
|x\ket \to H|x\ket = \frac{1}{\sqrt p}\sum_{y=0}^{p-1} (-1)^{x\otimes y}, 
|y\ket  
\end{equation}
where $\otimes$ means mod $p$ product. The Hadamard state $H|x\ket$ for a 
$p$-qubit $|x\ket$ is therefore a quantum superposition of 
all $p$-qubit states, labeled by $0\!\le\!y\!<p$, each taken with its 
parity with respect to $|x\ket$. 
For a sequence of $n$ p-qubits $|X\ket = |x_1,\ldots,x_n\ket$, therefore, 
the Hadamard transform is given by 
\begin{equation}
|X\ket \to H|X\ket = \frac{1}{\sqrt{p^n}}\sum_{y\in\omega} 
(-1)^{\bar x\cdot\bar y} |Y\ket,   
\end{equation} 
where $\bar x\cdot\bar y$ is bitwise dot product  mod p and the sum is taken 
over all $p^n$ sequences $|Y\ket$ consisting of $n$ p-qubits.    
\paragraph{Quantum Fourier transform.}
Starting from standard form of the Fourier 
transform of a vector in Hilbert space used in quantum computing 
\begin{equation}
F |x\ket = \sum_{y} |y\ket \exp(2\pi\imath \frac{x y}{\omega}) 
\label{ftn}
\end{equation}
which is used for the sequences of fixed length of $n$ $p$-qubits
(the total number of all this sequences is $\omega = p^n$), 
we can produce Fourier transform of any function 
\begin{equation}\displaystyle
\begin{array}{lcl}
F |\Psi_f\ket &=& F \sum_x |x\ket \otimes |f(x)\ket
= \sum_x F |x\ket \otimes |f(x)\ket \\
&=& \sum_{x,y} |y\ket \exp(2\pi\imath \frac{xy}{\omega}) \otimes 
|f(x)\ket = \sum_y |y\ket \otimes | g(y) \ket,
\end{array}
\end{equation} 
where $g(y)$ is the Fourier transform of $f(x)$.
Further applications of $p$-qubits to quantum computing, quantum 
database search, quantum cryptography \etc can be considered in a 
straightforward way following the known algorithms for traditional 
qubits ($p=2$).
\section{Conclusion}
The origin of space-time geometry from a set of relations between discrete 
objects by Big Bang or by other scenario is a challenging problem of 
quantum field theory and cosmology. The studies in this field give 
rise both to the development of new field theoretical methods related 
to quantum cosmology and to the development of new mathematical methods 
based on number theory, that can be used in different fields from 
information protection and coding theory to molecular biology. This 
paper, being partially based on physical intuition rather than 
rigorous axiomatics, makes in attempt to outline some multiscale methods based 
on on the $p$-adic generalization of the wavelet transform, the tool 
widely used 
in signal processing and data compression. We hope this research will 
be followed by more rigorous mathematical consideration of problems 
and algorithms outlined above in this paper.      

The other point of this paper was to attract attention to the possible 
generalization of quantum computing ideas to the quantum systems with 
$p>2$ orthogonal quantum states. Such systems, if organized in 
hierarchic structures, could be used for the hierarchic information 
storage. From mathematical point of view, the quantum states of those 
hierarchic structures can be discribed by wave functions depending on 
$p$-adic argument, and thus providing a new intriguing application of 
$p$-adic quantum mechanics.
\section*{Acknowledgement}
The work was partially supported by Russian Foundation for Basic Research,
Project 03-01-00657.
%\bibliographystyle{alpha}
%\bibliography{padic}

\end{document}